\documentclass[aps,prb,twocolumn,showpacs,preprintnumbers,amsmath,amssymb,superscriptaddress]{revtex4}
\usepackage{dcolumn}                    
\usepackage{bm}                        
\usepackage{graphicx}
\usepackage{times}
\usepackage{epstopdf}
\begin{document}

\newcommand{\e}[1]{\mathrm{e}^{#1}}

\newcommand{\vecr}{\boldsymbol{r}} 
\newcommand{\ve}{\boldsymbol{E}} 
\newcommand{\vv}{\boldsymbol{v}}
\newcommand{\vb}{\boldsymbol{B}}
\newcommand{\vs}{\boldsymbol{S}}
\newcommand{\vk}{\boldsymbol{k}}
\newcommand{\vn}{\boldsymbol{\hat{n}}}

\newcommand{\muS}{$\mu_{s}$}
\newcommand{\muN}{$\mu_{n}$}

\newcommand{\eq}{Eq.}
\newcommand{\eqs}{Eqs.}
\newcommand{\cf}{\textit{cf. }}
\newcommand{\ie}{\textit{i.e. }}
\newcommand{\eg}{\textit{e.g. }}
\newcommand{\etal}{\emph{et al.}}
\def\i{\mathrm{i}}    

\renewcommand{\section}[1]{{\par\it #1.---}\ignorespaces}

\title{Majorana fermions in spin-orbit coupled ferromagnetic Josephson junctions}
\author{Annica M. Black-Schaffer}
 \affiliation{Nordic Institute for Theoretical Physics (NORDITA), Roslagstullsbacken 23, S-106 91 Stockholm, Sweden}
 \affiliation{Department of Physics and Astronomy, Uppsala University, Box 516, S-751 20 Uppsala, Sweden}
 \author{Jacob Linder}
\affiliation{Department of Physics, Norwegian University of Science and Technology, N-7491 Trondheim, Norway}
\date{\today}

\begin{abstract}
We study all possible Majorana modes in two-dimensional spin-orbit coupled ferromagnetic  superconductor-normal state-superconductor (SNS) Josephson junctions and propose experiments to detect them. With the S region in a non-trivial topological phase and a superconducting phase difference $\phi = \pi$ across the junction, two delocalized Majorana fermions with no excitation gap appear in the N region.
In addition, if S and N belong to different topological phases and have well-separated the Fermi surfaces, localized Majorana fermions with a finite excitation gap also emerge at both SN interfaces for all $\phi$.
\end{abstract}
\maketitle
The search for Majorana fermions in condensed matter physics has recently escalated \cite{Wilczek09, Franz10}. A Majorana fermion is its own anti-particle and not only would its discovery be an extraordinary achievement, but it also supports fault-tolerant topological quantum computation in two dimensions (2D), where the qubits are decoherence free and protected against local perturbations by topology \cite{Kitaev03}. This is  due to the Majorana fermion obeying non-Abelian statistics, where exchange operations between particles do not just produce overall phase factors, as in the case of fermions or bosons, but are non-commutative \cite{Read00, Nayak08}.
Traditionally, the $\nu =5/2$ state in quantum Hall systems \cite{Read00} and spinless (spin-polarized) $p_x+ip_y$ superconductors, proposed e.g.~in stronitium ruthenate \cite{DasSarma06} and cold atom systems \cite{Gurarie05}, have been considered for finding Majorana fermions. More recently, also strongly spin-orbit coupled (SOC) systems with a spin-singlet $s$-wave superconducting order parameter have been shown to host Majorana fermions in the presence of a magnetic field, due to an effective $p+ip$-wave symmetry \cite{Fu08, Stanescu10, Black-Schaffer11QSHI,Alicea10}. This includes both topological insulators (TIs) \cite{Fu08, Fu09,Linder10PRL} and more generic SOC semiconductors \cite{Sau10, Alicea10, Lutchyn10, Sato09, Sato10, Oreg10, Linder10}. Especially SOC semiconductors have generated much attention, both due to technological maturity and to the experimentally demonstrated superconducting proximity effect in InAs \cite{Chrestin97}, and will be the focus here.
Majorana fermions have been shown to appear in these SOC materials at vortex cores and at external edges, i.e.~edges to the vacuum \cite{Sau10,Lutchyn10,Sato09,Sato10,Oreg10,Linder10}. However, in order to utilize the Majorana fermions it will be necessary to contact the structure producing them and therefore it is the "internal" edge created at a superconducting-normal state (SN) interface that is, by far, the most interesting.
In this Rapid Communcation we will answer the question of when Majorana fermions appear at internal edges in a generic 2D SOC semiconductor and how they can be detected. We will do this by studying, both analytically and numerically, a finite length SNS Josephson junction in a Rashba SOC 2D system. 
At first glance one could possibly expect a SN interface to behave the same way as a S-vacuum edge. 
However, due to the superconducting proximity effect, the effective superconducting gap $\Delta_{\rm eff}$ is small, but finite, also on the N side of the interface. Consequently, the topological phase (TP) in the N region is determined by setting $\Delta =\Delta_{\rm eff} \sim 0$ in the same phase diagram as used for the S region, which can produce a significantly different result from a S-vacuum interface. We will here show the following:
(1) If S and N belong to the same non-trivial TP, there are two zero-energy Majorana fermions when the superconducting phase difference across the junction  is $\phi = \pi$, due to closing of the Andreev bound state (ABS) spectrum in the junction. This has already been established in $L\rightarrow 0$ junctions \cite{Sau10}, but here we also show that these Majorana modes persist for any $L$ and, moreover, that they are delocalized over the whole N region and that there is no gap to normal fermionic excitations. However, since the Majorana modes have different fermion parity, they can be detected through a $4\pi$ contribution to the Josephson current. We call these {\it ABS Majorana modes}.
(2) If S and N belong to different TPs, of which S is in a non-trivial phase, there will be one chiral Majorana mode localized at each SN interface for all $\phi$. We call these {\it phase boundary (PB) Majorana modes}. However, if the Fermi surfaces (FSs) of S and N are not well-separated, the ever-present ABS spectrum in the junction will, due to hybridization, destroy the PB Majorana modes.
Since the PB Majorana modes are well-localized and have a finite excitation gap, they can both be detected by a local density of states (LDOS) probe, such as scanning tunneling microscopy, and are the only Majorana mode which qualify for quantum computation.
%
%
\section{Model}
We use a 2D square lattice model with nearest neighbor hopping $t = 1$, doping $\mu <-2t$, and Rashba SOC $\alpha$ to model a generic 2D SOC semiconductor: $H_{kin} = -t\sum_{\langle i,j\rangle,\sigma} c_{i \sigma}^\dagger c_{j,\sigma} - \sum_{i,\sigma} \mu(i) c_{i \sigma}^\dagger c_{i,\sigma} +\alpha \sum_i [(c_{i\uparrow}^\dagger c_{i+x \downarrow} - c_{i \downarrow}^\dagger c_{i+x \uparrow}) - \i(c_{i\uparrow}^\dagger c_{i+y \downarrow} + c_{i \downarrow}^\dagger c_{i+y \uparrow}) + {\rm H.c.}]$. Here $c_{i\sigma}$ is the fermion annihilation operator on site $i$ with spin $\sigma$. In order to break the spin-degeneracy and produce Majorana fermions we add a Zeeman field to the whole SNS structure: $H_{V_z} = -V_z \sum_{i \sigma \sigma'} \sigma^z_{\sigma \sigma'} c_{i\sigma}^\dagger c_{i\sigma'}$. Experimentally $V_z$ can be provided by proximity to a ferromagnetic insulator. Finally, in the S regions of the structure we model the proximity induced superconducting state by $H_{\Delta} = \sum_i \Delta(i)c_{i\uparrow}^\dagger c_{i\downarrow}^\dagger + {\rm H.c.}$. The superconducting order parameter is $\Delta = \Delta_0$ in the S regions and zero otherwise. 
We also assume smooth interfaces and Fourier transform with momentum $k_y$ in the direction parallel to the SN interfaces. 
Our model is equivalent to that of Sato \etal\ \cite{Sato10} and we quickly restate the four different TPs possible for $\mu<-2t$: (I): $0<V_z^2<(4t+\mu)^2+\Delta^2$, (II): $(4t+\mu)^2+\Delta^2<V_z^2<\mu^2+\Delta^2$, (III): $\mu^2+\Delta^2<V_z^2<(4t-\mu)^2+\Delta^2$, and (IV): $(4t-\mu)^2+\Delta^2<V_z^2$. 
It is important to know here that phase (I) is a trivial TP with two (non-superconducting) FSs centered around $\Gamma = (0,0)$, (II) is a non-trivial TP with one FS centered around $\Gamma$, (III) is also a non-trivial TP but with its FS centered around $M=(\pi,\pi)$, whereas (IV) is a trivial band insulator. Thus, Majorana modes exist at S-vacuum edges at $k_y = 0$ in phase (II) and at $k_y = \pi$ in phase (III) \cite{Sato10}. Note that due to the superconducting proximity effect at SN interfaces, $\Delta$ in the above PB equations needs to be the {\it effective} order parameter and should thus be $\Delta_0$ for the S regions and $\Delta_{\rm eff} \sim 0$ for the N region. 

%
\section{ABS Majorana modes}
We first consider an analytical description of the above model in phase (II) at very low carrier concentrations, i.e.~an effective model of a lightly doped semiconductor with one FS centered around $\Gamma$ in both S and N. In this regime we can approximate the band structure as $\mathcal{E}_{\vk} = |\vk|^2/(2m) - \mu' - \sqrt{\alpha'^2 |\vk|^2+V_z^2}$, where $m = (2ta^2)^{-1}$, $\mu' = \mu+4t$, $\alpha' = a\alpha$, with $a = 1$ being the lattice constant, but for brevity we will here drop the prime $(')$. This directly connects to previous continuum model work on SOC semiconductors \cite{Sau10, Alicea10} and will also allow us to analytically extract the ABS spectrum.
The $s$-wave spin-singlet order parameter $\Delta_0$ has the component $\Delta_{\vk} = -\alpha\Delta_0(k_y-\i k_x)/(2\sqrt{\alpha^2|\vk|^2+V_z^2})$ in this band after a pseudospin-transformation. Effectively, we are thus left with a Hamiltonian which maps perfectly onto a spinless $p+ip$ superconductor. 
To model a realistic scenario, we allow for both different chemical potentials and masses of the quasiparticles in the S and N parts of the junction, denoted $\{\mu_s, m_s\}$ and $\{\mu_n,m_n\}$, respectively. In this way, an effective resistance is present upon transmission between the S and N regions due to the Fermi-vector mismatch. 
In order to compute the ABS energies, we set up the wavefunctions in each of the three regions in the junction and match them appropriately at the interfaces, using a framework similar to Ref.~\cite{Linder10}. 
Lengthy calculations provide the allowed ABS energies $\pm \varepsilon_n$ where
%
\begin{align}
\label{eq:abs}
\varepsilon_{n} &= \frac{\alpha\Delta_0}{\sqrt{2}}\Big([(q_x^2-k_x^2)^2(1-\mathcal{C})\cos^2(\gamma/2) + 8q_x^2k_x^2\notag\\
&\times\cos^2(\phi/2)]/(\alpha^2k_F^2+V_z^2)[2k_x^4+2q_x^4+12q_x^2k_x^2 \notag\\
&- 2(k_x^2-q_x^2)\mathcal{C}]\Big)^{1/2} \text{ with } \mathcal{C} = \cos2q_xL.
\end{align}
Above, $k_x = \sqrt{k_F^2 - k_n^2}$ where $k_F = [2m_s\mu_s + 2m_s^2\alpha^2 + 2m_s\sqrt{V_z^2+m_s^2\alpha^4 + 2m_s\mu_s\alpha^2}]^{1/2}$ while $q_x = (m_s/m_n)\sqrt{q_F^2 - k_n^2}$ with $q_F = [2m_n\mu_n + 2m_n^2\alpha^2 + 2m_n\sqrt{V_z^2+m_n^2\alpha^4 + 2m_n\mu_n\alpha^2}]^{1/2}$. Here, $k_y = k_n$ is the transverse momentum index which is quantized for a finite width $W$ of the junction, whereas the phase factor $\gamma$, defined via $\e{\i\gamma}=(k_y+\i k_x)/k_F$, is related to the $k$-space structure of the superconducting gap. The above bound-state expression can then be used to calculate the Josephson current in the short-junction regime $L\ll\xi$: $I(\phi) = \frac{2e\Delta_0}{\hbar} \sum_n \frac{\partial \varepsilon_{n}}{\partial \phi} t_n$ with $t_n\equiv\tanh(\beta\varepsilon_{n}/2)$ and $\beta$ the inverse temperature. As a consistency check for the above result, we briefly consider the limit of equal masses and chemical potentials in the system, $m_s=m_n$ and $\mu_s=\mu_n$. In this case, Eq.~(\ref{eq:abs}) is seen to reduce to the form $\varepsilon_{n} = \sqrt{D}\cos(\phi/2)$ where $D$ is a transmission constant independent on $L$. This is in agreement with Ref.~\cite{Kwon04}, where 1D tunneling between pure $p$-wave superconductors was considered. 

%
\begin{figure}[hbt!]
\centering
\resizebox{0.5\textwidth}{!}{
\includegraphics{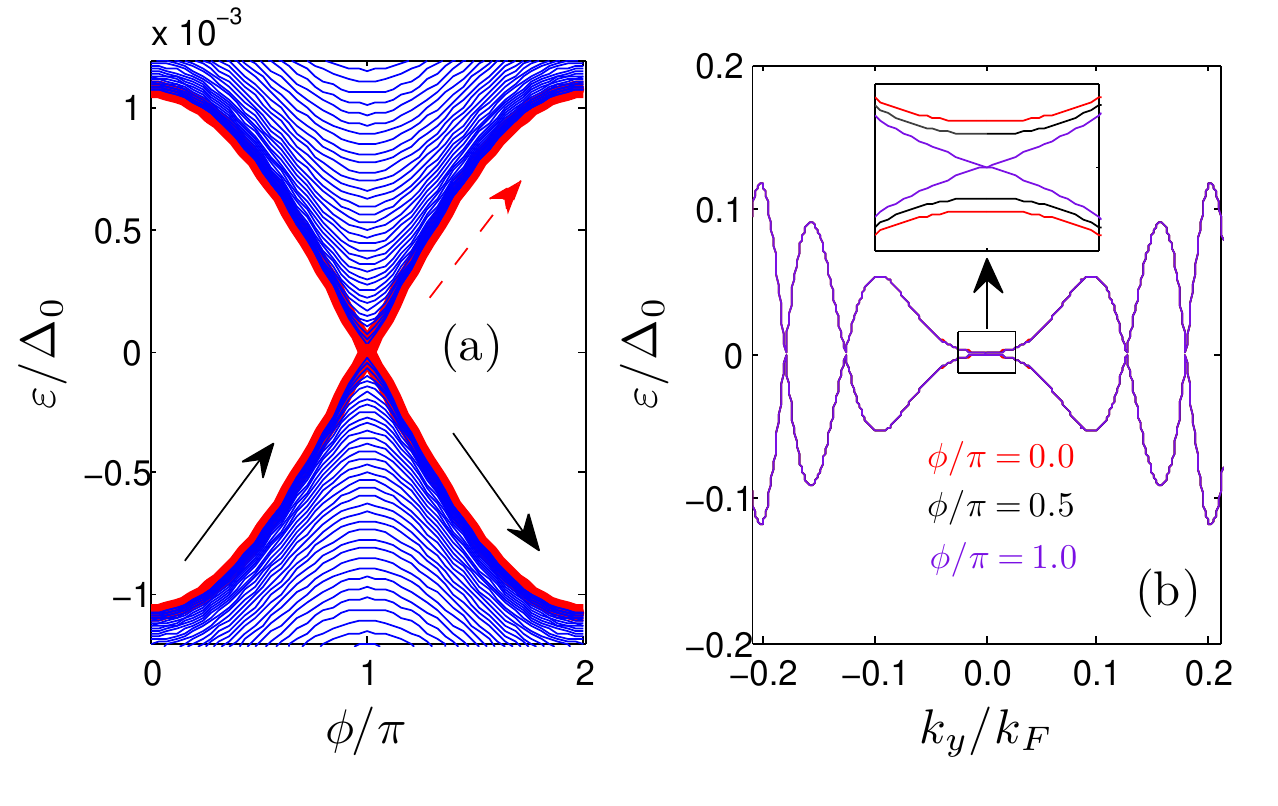}}
\caption{(Color online) (a) ABS energy spectrum in the Josephson junction as a function of $\phi$. Blue (thin) lines correspond to different finite $k_y$, whereas the red (thick) line denotes the mode $k_y=0$. In an equilibrium situation, the lower energy states $\varepsilon<0$ would be populated upon increasing $\phi$ (black solid arrows). For an adiabatic ac voltage bias it is possible to keep one single branch populated (red dashed arrow). (b) ABS energy spectrum as a function of $k_y$. The inset shows a zoom-in of the behavior near $k_y=0$. The parameters are chosen to be relevant to experiments but still allow the one-band model: $\Delta_0 = 0.02$~meV (a fraction of the bulk superconducting gap), $V_z = 1$~meV (due to proximity to a ferromagnetic insulator), $m_n/m_e=0.04$ and $m_n\alpha^2 = 0.01$~meV (pertaining to InAs quantum wells\cite{Heida98}), $m_s/m_e=0.2$ and $m_s\alpha^2=0.05$~meV (motivated by the proximity to bulk superconducting leads), $\mu_s=\mu_n=0.1$ meV (tunable via an overall gate voltage), and $W/\xi =10$.}
\label{fig:energyABS} 
\end{figure}
In Fig.~\ref{fig:energyABS} we display the energy spectrum of the ABSs in the junction. In Fig.~\ref{fig:energyABS}(a) we show energy versus $\phi$ for multiple transverse modes $k_y$. The most striking feature is the appearance of a zero-energy crossing precisely at $\phi=\pi$ for $k_y = 0$. 
For finite values of $k_y$, the ABS-levels repel each other just as in ordinary $s$-wave superconductors. In Fig.~\ref{fig:energyABS}(b), we consider energy versus $k_y$. The ABS-levels oscillate strongly with momentum, similarly to the numerical lattice-results discussed below. Again, we see the zero-energy crossing at $k_y=0$ when $\phi = \pi$. We find both here and in the numerical results that the two states associated with the zero-energy crossing, the two ABS Majorana fermions, are fully delocalized in the N region. However, as pointed out in Refs.~\cite{Fu09, Lutchyn10}, the absence of processes that violate the conservation of fermion parity locally ensures that there is no transition between these two states, thus keeping their Majorana nature intact.
The question is now, is it possible to experimentally identify the above mentioned ABS Majorana modes? Here we demonstrate a route for doing so via standard transport measurements. 
Consider Eq.~(\ref{eq:abs}) and the normally incident mode $n=0$ for which we obtain: $\varepsilon_{0} = \sqrt{D(L)}\cos(\phi/2)$. Note that the effective transmission coefficient is now dependent on the length $L$ of the junction. The $4\pi$-periodicity is given by the zero-energy levels crossing at $\phi = \pi$ instead of repelling each other.
Since the Josephson current may be written as $I(\phi) = \frac{2e\Delta_0}{\hbar}[\frac{\partial\varepsilon_{0}}{\partial \phi}t_0 + \sum_{n\neq0} \frac{\partial \varepsilon_{n}}{\partial \phi}t_n]$, it is clear that as long as the first term has an appreciable magnitude compared the higher $n\neq0$ modes (which is expected since the transmission probability peaks at normal incidence), its periodicity should be reflected in the current-phase relation of the total Josephson current. It is important to note that the observation of this 4$\pi$-periodicity, or equivalently a fractional Josephson effect, is not possible for a dc bias under equilibrium conditions. The reason for this is that then the population of the positive and negative energy branches of the ABSs in Fig.~\ref{fig:energyABS} would always be zero and filled, respectively. The negative of these energy branches is 2$\pi$-periodic, and thus the effect would be lost. On the other hand, by applying a bias voltage to the junction with a Josephson period of $\tau_J$, one can ensure to populate a single branch when adiabatically varying $\phi(t)$ as long as $\tau_J$ is smaller than the relaxation time associated with e.g.~a bound-state emitting a photon and then relaxing into the negative branch. By keeping only one single branch populated for all phases in this way, one would observe precisely the announced 4$\pi$-periodicity, which has previously been discussed for 1D $p+ip$ systems \cite{Kwon04, Fu09, Lutchyn10}. The experimental technique for non-equilibrium population of a single branch in a controllable fashion has been clearly demonstrated in the context of Josephson junctions \cite{Baselmans02}.

%
\section{PB Majorana modes}
\begin{figure}[tb!]
\includegraphics[scale = 0.98]{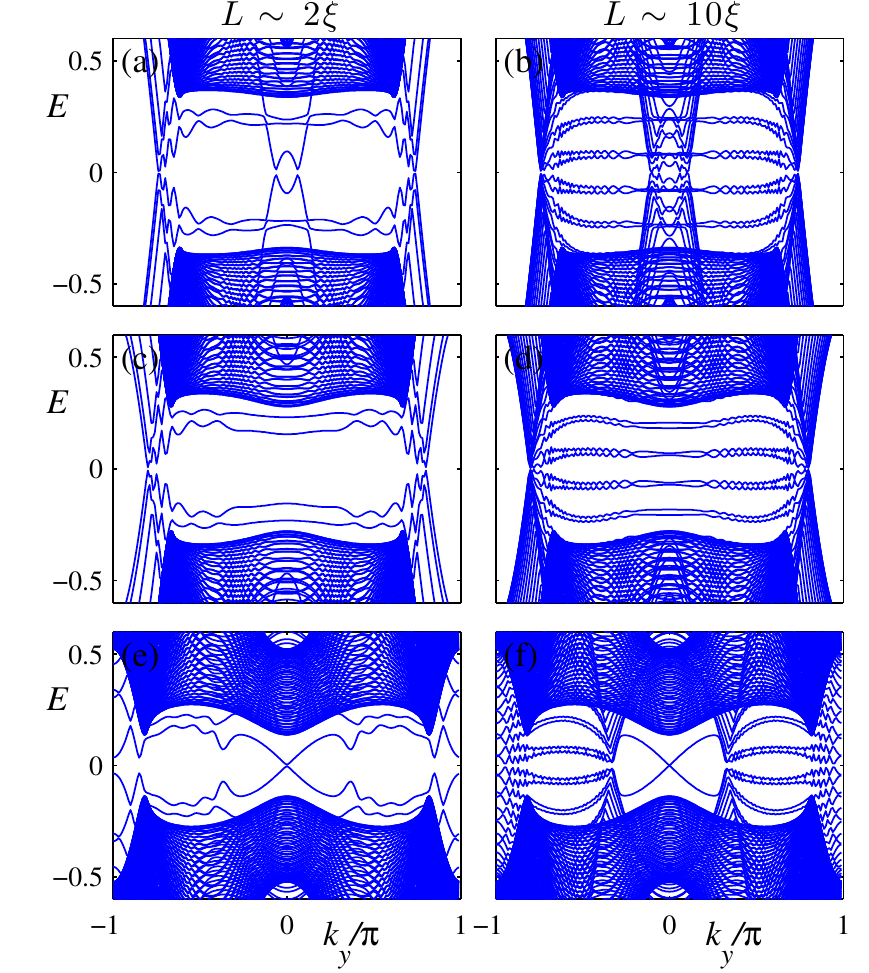}
\caption{\label{fig:bands} (Color online) Eigenvalue spectrum for $\phi = 0$ junctions with $L =8 \sim 2 \xi$ (a),(c),(e) and $L = 40$ (b),(d),(f) in the case where S belongs to phase (II) and N to phase (I) with $V_z = 1.2$ (a),(b), (II) with $V_z = 1.8$ (c),(d), and (III) with $V_z = 3$ (e),(f). Here $\alpha = 1$, $\Delta_0 = 0.4$, \muS = -3.5, and \muN = -2.5. For simplicity only $V_z$ was varied to produce the different TPs in N while keeping a fixed Fermi level mismatch in the form of $\mu_s \neq \mu_n$, but the different TPs can be implemented using a wide range of parameters.
}
\end{figure}

Having established the appearance of Majorana fermions at $\phi=\pi$ when S and N belong to the same non-trivial TP, we return to the lattice model and its more general phase diagram. Here we will show that a Majorana fermion can also be located at the SN interface for any $\phi$ if, in addition to S being in a non-trivial TP, N is in {\it another} phase. The position of this PB Majorana fermion will be at the PB between the S and the N region and is thus a manifestation of the termination of the non-trivial TP in S, akin to a S-vacuum edge. Therefore the inverse proximity effect,  i.e.~the reduction of $\Delta$ on the S side of the interface captured in a self-consistent treatment, can move  the Majorana fermion well into the S region. 
Below we only report standard non-self-consistent results, but we have confirmed our results even when including self-consistency.
In Fig.~\ref{fig:bands} we plot the eigenvalue spectrum at $\phi = 0$ for different junction lengths [the superconducting coherence length is defined as $\xi = 2\hbar v_F/(\pi \Delta_0)$], for S in phase (II) and N in phases (I)-(III), counted from the top. 
%
%
The case treated analytically above, with S and N in (II), corresponds to Figs.~\ref{fig:bands}(c) and \ref{fig:bands}(d) but here $\phi=0$, so no ABS Majorana fermions are present. We see that the ABS spectrum is somewhat more spread out in $k_y$ in the numerical solution, due to the anharmonicity of the band structure, and that more and lower lying ABS levels appear for longer junctions. However, we reproduce all significant results from the analytical treatment. Most notably, the zero-energy ABS Majorana fermions appear only at $\phi = \pi$ and $k_y =0$. 
If instead S and N are both in phase (III), the results are analogous with the exception of the Majorana fermions now appearing at $k_y = \pi$. This is a direct consequence of the ABS spectrum being centered around the same $k_y$ value as the FS in N.
Moving on to Figs.~\ref{fig:bands}(e) and \ref{fig:bands}(f), where N is in phase (III), the ABS spectrum is centered around $k_y = \pi$ and it is again rather spread out in $k$-space. Here we also have a proper PB between the S and the N regions with the Majorana modes associated with phase (II) in S located at $k_y = 0$. Thus, the ABS spectrum and the PB Majorana modes are well separated in $k$-space, as clearly seen in Figs.~\ref{fig:bands}(e) and \ref{fig:bands}(f), and the PB Majorana mode is present.
Finally in  Figs.~\ref{fig:bands}(a) and \ref{fig:bands}(b), N is in phase (I) and there is again a PB between S and N. However, both the ABS spectrum and the PB Majorana modes are now centered at the same $k_y =0$ point, and we see a large hybridization between these two features, resulting in the destruction of the zero-energy PB Majorana fermions. Note that this hybridization is not only strong at large $L$, where the ABS spectrum appears at low (but finite) energies, but exist even in Fig.~\ref{fig:bands}(a), where the ABSs are located at energies $\sim 0.2t$ for small $k_y$-values. In the very short junction limit, where the ABS spectrum joins the bulk continuum, we start seeing small remanent traces of the PB Majorana modes at the SN interfaces, although in our lattice model $L$ is now so short that the Majorana modes themselves start to significantly overlap, causing a finite gap. It is thus hard to determine which effect is largest in terms of gapping out and destroying the two distinct interface Majorana modes in the $L\rightarrow 0$ case.
Not shown in Fig.~\ref{fig:bands} is N in phase (IV). Here N is a band insulator which cannot support ABS levels inside the insulating gap, and we thus find PB Majorana modes at $k_y =0$ for S in phase (II). 
Lastly, analogously to the results in Fig.~\ref{fig:bands}, for S in phase (III) we find PB Majorana modes at $k_y = \pi$ for N in phases (I),(II), and (IV), as then there is a PB between S and N and the ABS spectrum does not interfere.
We thus conclude that PB Majorana modes exist only when S is in a non-trivial TP, there is a PB between S and N and, most importantly, only when the FSs of S and N are {\it not} centered at the same $k$-point.

%
\begin{figure}[t]
\includegraphics[scale = 0.98]{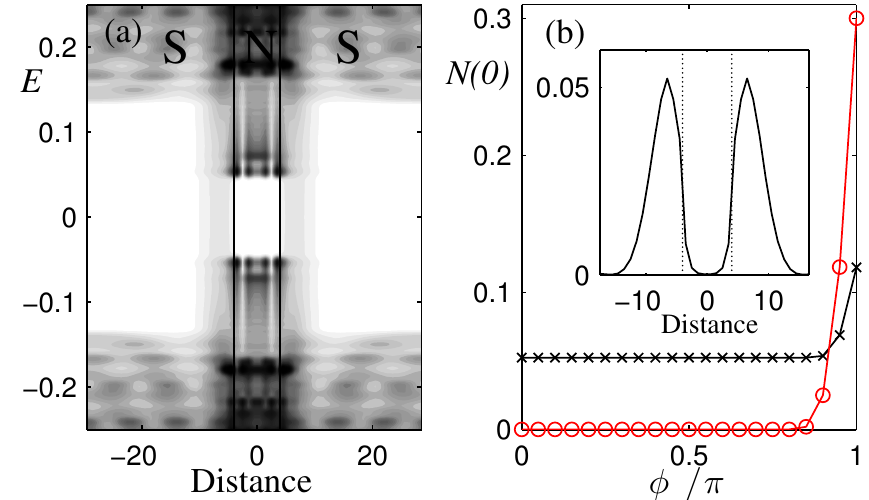}
\caption{\label{fig:LDOS} (Color online) (a) LDOS  per unit cell and energy (white = 0 to black = 0.55) as a function of position from the junction center (in units of $a$) and energy for $\phi = 0$. The vertical lines mark the SN interfaces. (b) $N(0)$ (integrated LDOS for $|E|<0.01$) as a function of $\phi$ at Majorana peak position (black, $\times$) and in the middle of N (red, $\circ$). The inset displays the two interface Majorana fermions by plotting $N(0)$ for $\phi =0$ as a function of distance. The dotted vertical lines mark the SN interfaces. Here $\alpha = 1$, $\Delta_0 = 1$, $V_z = 3.5$, \muS = -3.5, \muN = -2.5, and $L = 8 \sim 2\xi$.
}
\end{figure}
In Fig.~\ref{fig:LDOS} we explore the spatial distribution of the PB Majorana modes in the prototype case of S in phase (II) and N in phase (III) [cf.~Fig.~\ref{fig:bands}(e)]. 
In Fig.~\ref{fig:LDOS}(a) the LDOS as a function of both energy and distance from the middle of the junction is plotted. The bulk gap appears at $\sim 0.12$ in S, whereas the ABS spectrum in N reaches down in energy to at $\sim 0.05$ for $\phi = 0$. The light gray band at even low energies on either side of the interfaces are the PB Majorana modes, the constant DOS being a feature of the Dirac spectrum in 1D. 
To complement this data we plot in the inset in Fig.~\ref{fig:LDOS}(b) $N(0)$, the low-energy carrier density, as a function of distance, which shows the well resolved PB Majorana fermions at the two SN interfaces. Thus the PB Majorana modes constitute two counter-propagating chiral Majorana modes well localized to the two SN interfaces.
Finally in the main panel in Fig.~\ref{fig:LDOS}(b) we explore the dependence on the superconducting phase difference $\phi$ across the junction, by plotting $N(0)$ at the peak position of the Majorana fermion (black crosses) and in the middle of the junction where the ABS spectrum is present (red circles). Here we see that the Majorana fermion persists for all $\phi$. The slight increase in $N(0)$ at the Majorana position at very large $\phi$ is due to the ABS spectrum finally closing at $\phi = \pi$, as seen clearly in the red curve with circles. Note, however, that while the PB Majorana modes are only well resolved in energy for $\phi < \pi$ in $k$-integrated data, they are still well resolved in $k$-space for all $\phi$. This is in contrast to the ABS Majorana mode which, first of all, only exists at zero energy for $\phi = \pi$, and even then, they are not separated in energy from other low-lying parts of the ABS spectrum.
The localization to the SN interfaces, and the existence at all $\phi$, which allows for energy separation from the ABS spectrum, means that the PB Majorana modes can be easily detected by a LDOS probe such as scanning tunneling microscopy. To allow for a broad window of detection, short junctions are preferred as then the ABS spectrum appears at higher energies. 


\section{Acknowledgments}
A.M.B.-S. thanks Eddy Ardonne and Hans Hansson for valuable discussions and the Swedish Research Council (VR) for financial support.


\end{document}